\RequirePackage{fix-cm}
\documentclass[smallextended]{svjour3}       
\smartqed  
\usepackage{graphicx}
\usepackage{multirow}
\usepackage{epsfig}
\journalname{}

\title{Universal violation of pentagon inequalities in four-state systems}

\author{Mordecai Waegell and P.K. Aravind }

\authorrunning{M.Waegell, P.K. Aravind}

\institute{M.Waegell, P.K. Aravind \at
Physics Department, Worcester Polytechnic Institute, Worcester, MA 01609, U.S.A.\\
\email{caiw@wpi.edu, paravind@wpi.edu}}

\date{\today}

\begin{document}
\maketitle
\begin{abstract}
The 60 real vectors derived from the vertices of a 600-cell are shown to yield a number of pentagon inequalities that are satisfied by realistic noncontextual theories but violated by quantum mechanics. The replicas of these inequalities cover Hilbert space so densely that every real four-dimensional vector violates at least one of them. It is pointed out that this set of 60 vectors contains numerous ``N-gons" (generalizations of pentagons) that may be of interest in connection with demonstrations of contextuality.
\end{abstract}

\section{\label{sec:Intro}Introduction}

Specker \cite{Specker} pointed out many years ago that the transitivity of implication does not always hold for the outcomes of a sequence of quantum measurements. This observation has been expanded on in a number of ways in recent years. Klyachko et al \cite{Klyachko} derived an inequality that must be satisfied by any realistic noncontextual theory for the averages of certain measurements carried out on a qutrit. If the measurements are associated with the vertices of a pentagon, then violations of the inequality can be interpreted as a failure of the transitivity of implication between the outcomes of neighboring measurements on the pentagon. For this reason, the Klyachko inequality can be referred to as a ``pentagon inequality". The Klyachko inequality can be generalized by replacing the qutrit by a qudit (i.e. a d-state system with d $>$ 3) and by considering measurements associated with graphs more complicated than a pentagon. These generalizations, and their implications for nonlocality and other issues in quantum information, have been explored by a number of authors \cite{Bengtsson,Liang,Severini,Cunha,Aolita,Sadiq}. In parallel with these theoretical investigations, violations of the pentagon and related inequalities have been observed in a variety of four-state systems realized using ions \cite{Kirchmair}, neutrons \cite{Bartosik}, photons \cite{Amselem} and nuclear spins \cite{Moussa}.\\

The purpose of this paper is to present a number of pentagon inequalities for a four-state system using a system of 60 real rays in four dimensions derived from the vertices of a 600-cell. We \cite{Waegell2011} recently used these rays to give a large number of ``parity proofs" of the Kochen-Specker (KS) theorem \cite{KS1967}. Here we show that these same rays yield a number of pentagon inequalities that can be used to rule out realistic noncontextual theories. This in itself may not cause much surprise since violations of such inequalities have been exhibited earlier. However what makes our results interesting is that they permit us to give an affirmative answer to the following question raised at the end of \cite{Bengtsson}: ``Take any Kochen-Specker graph in 3 or 4 dimensions leading to an absolute rather than a probabilistic Kochen-Specker contradiction. Enumerate all its (pentagon) subgraphs. Is it by any chance the case that every real vector in Hilbert space will violate at least one of the corresponding (pentagon) inequalities?" We show that the 60 rays of the 600-cell yield a large number of pentagon inequalities that cover Hilbert space so densely that every real four dimensional vector leads to a violation of at least one of these inequalities. This is what we mean when we state, in the title to this note, that the violation of the pentagon inequalities is universal. We also discuss violations of heptagon inequalities by the same set of 60 rays, as well as some related geometrical matters that may be of interest in connection with discussions of contextuality.

\section{\label{sec:Penta1} The Klyachko ``pentagon" inequality}

Let $|\psi_{k}\rangle, (k \in 0,1,2,3,4)$, be five states in a four-dimensional Hilbert space whose only orthogonalities are given by  $\langle\psi_{k}|\psi_{k\oplus1}\rangle = 0$, where $\oplus$ denotes addition modulo 5. These states can be placed at the vertices of a pentagon in such a way that the only orthogonalities between them are between states at adjacent vertices. For this reason, we will refer to these states as a ``pentagon" and the inequality satisfied by them, to be derived now, as the ``pentagon inequality". Consider the operator $\Sigma = \sum_{i = 0} ^{4} |\psi_{k}\rangle\langle \psi_{k}|$, which we will term the ``pentagon operator". According to noncontextual realism, $\langle\Sigma\rangle \leq 2$, with the average taken in an arbitrary quantum state. This conclusion follows from the facts that (1) any projector, $|\psi_{k}\rangle \langle \psi_{k}|$, returns the value 0 or 1 upon measurement, and (2) no two orthogonal projectors both return a 1 when measured on the same state, if one makes the additional assumption, characteristic of noncontextual realism, that all projectors possess definite values in any quantum state that merely reveal themselves upon measurement\cite{noncon}. These conditions require one to assign a 0 or 1 to each of the projectors in a pentagon operator in such a way that no two neighboring projectors are both assigned a 1. But this implies that no more than two of the projectors can be assigned a 1 for a given quantum state, and hence that the average value of the pentagon operator in any state can never exceed 2. This is the pentagon inequality. In order to rule out noncontextual realism, all one has to do is to find an example of a quantum state in which the expectation value of the pentagon operator is greater than 2. This can be done by finding a ``pentagon" of states whose pentagon operator has a largest eigenvalue greater than 2. The eigenstate corresponding to this largest eigenvalue is then an example of a state in which the average value of the pentagon operator can be measured to reveal a conflict with noncontextual realism (this choice in fact leads to the largest conflict possible). We will refer to a pentagon of states whose pentagon operator has an eigenvalue greater than 2 as a ``conflict pentagon".\\

The pentagon inequality was first derived for a three-state system by Klyachko et al \cite{Klyachko}, who also identified a set of states for which the pentagon operator had a largest eigenvalue of $\surd5 = 2.236$ (which is appreciably above the bound of 2 for a realist noncontextual theory). Pentagons in four dimensions were investigated by Badzi\c{a}g et al \cite{Bengtsson}, who found a number of different examples of conflict pentagons. One of them was contained in a set of 18 vectors that had been used earlier \cite{Cabello1996} to prove the KS theorem, and the corresponding pentagon operator had a largest eigenvalue of 2.171. This result suggested two ideas to us. The first was to examine other sets of Kochen-Specker vectors in four dimensions to see if they contained conflict pentagons. However we found that a Kochen-Specker set of vectors need not, in general, contain any conflict pentagons in it. The second was to look at the set of 24 vectors of Peres \cite{Peres1991} that contained the 18 just mentioned, and to see whether the several conflict pentagons contained in this larger set led to a violation of at least one of the corresponding pentagon inequalities by an arbitrary real vector. However this hope was dashed by the discovery that the 24 defining vectors themselves led to an expectation value of no larger than 2 for any of the pentagon operators in the set. Despite these initial setbacks, we succeeded in identifying a symmetrical set of 60 vectors in four dimensions that allowed us to succeed at both of the tasks just mentioned. It is to this set that we now turn.

\section{\label{sec:Penta2} Pentagon inequalities based on the 600-cell}

The system of 60 states we use in this work is associated with the vertices of a four-dimensional regular polytope known as the 600-cell. We pointed out recently \cite{Waegell2011} that these states yield over a 100 million parity proofs of the KS theorem. Each of the 60 states is associated with an antipodal pair of vertices of the 600-cell, and the (real) vectors that describe them are shown in Table \ref{tab:Ray}.\\

\begin{table}[htp]
\vskip-12pt
\addtolength{\tabcolsep}{10pt}
\centering 
\begin{tabular}{|c c c c|} 
\hline 
1 = $2 0 0 0 $ & 2 = $0 2 0 0 $ & 3 = $0 0 2 0 $ & 4 = $0 0 0 2 $ \\
5 = $1 1 1 1 $ & 6 = $1 1 \overline{1} \overline{1} $ &
7 = $1 \overline{1} 1 \overline{1} $ & 8 = $1 \overline{1} \overline{1} 1 $ \\
9 = $1 \overline{1} \overline{1} \overline{1} $ &
10 = $1 \overline{1} 1 1 $ & 11 = $1 1 \overline{1} 1 $ &
12 = $1 1 1 \overline{1} $ \\
\hline
13 = $\kappa 0 \overline{\tau} \overline{1} $ &
14 = $0 \kappa 1 \overline{\tau} $ &
15 = $\tau \overline{1} \kappa 0 $ & 16 = $1 \tau 0 \kappa $ \\
17 = $\tau \kappa 0 \overline{1} $ & 18 = $1 0 \kappa \tau $ &
19 = $\kappa \overline{\tau} \overline{1} 0 $ &
20 = $0 1 \overline{\tau} \kappa $ \\
21 = $1 \kappa \tau 0 $ & 22 = $\tau 0 \overline{1} \kappa $ &
23 = $0 \tau \overline{\kappa} \overline{1} $ &
24 = $\kappa \overline{1} 0 \overline{\tau} $ \\
\hline
25 = $\tau 0 1 \kappa $ & 26 = $0 \tau \overline{\kappa} 1 $ &
27 = $1 \overline{\kappa} \overline{\tau} 0 $ &
28 = $\kappa 1 0 \overline{\tau} $ \\
29 = $0 \kappa 1 \tau $ & 30 = $\tau 1 \overline{\kappa} 0 $ &
31 = $\kappa 0 \tau \overline{1} $ & 32 = $1 \overline{\tau} 0 \kappa $ \\
33 = $\tau \overline{\kappa} 0 \overline{1} $ &
34 = $0 1 \overline{\tau} \overline{\kappa} $ &
35 = $1 0 \overline{\kappa} \tau $ & 36 = $\kappa \tau 1 0 $ \\
\hline
37 = $\tau 0 \overline{1} \overline{\kappa} $ &
38 = $0 \tau \kappa \overline{1} $ & 39 = $1 \overline{\kappa} \tau 0 $ &
40 = $\kappa 1 0 \tau $ \\
41 = $\tau 1 \kappa 0 $ & 42 = $0 \kappa \overline{1} \overline{\tau} $ &
43 = $1 \overline{\tau} 0 \overline{\kappa} $ &
44 = $\kappa 0 \overline{\tau} 1 $ \\
45 = $0 1 \tau \kappa $ & 46 = $\tau \overline{\kappa} 0 1 $ &
47 = $\kappa \tau \overline{1} 0 $ & 48 = $1 0 \kappa \overline{\tau} $ \\
\hline
49 = $\kappa 0 \tau 1 $ & 50 = $0 \kappa \overline{1} \tau $ &
51 = $\tau \overline{1} \overline{\kappa} 0 $ &
52 = $1 \tau 0 \overline{\kappa} $ \\
53 = $1 0 \overline{\kappa} \overline{\tau} $ &
54 = $\tau \kappa 0 1 $ & 55 = $0 1 \tau \overline{\kappa} $ &
56 = $\kappa \overline{\tau} 1 0 $ \\
57 = $\tau 0 1 \overline{\kappa} $ &
58 = $1 \kappa \overline{\tau} 0 $ &
59 = $\kappa \overline{1} 0 \tau $ & 60 = $0 \tau \kappa 1 $ \\
\hline 
\end{tabular}
\caption{The 60 states associated with the vertices of a 600-cell. The states are unnormalized and represented as real, four-dimensional vectors of length 2. The four numbers following each state are its components in an orthonormal basis, with $\tau=(1+\surd5)/2$, $\kappa=1/\tau$, a bar over a number indicating its negative and commas omitted between components. The 60 vectors shown here represent half the vertices of a 600-cell centered at the origin, with their negatives representing the remaining vertices.}
\vskip-15pt
\label{tab:Ray} 
\end{table}

The 60 states can be separated, in 120 ways, into two sets of 30 states that each yield a parity proof of the KS theorem. Our new discovery is that each of the sets of 30 states that yields a parity proof contains several types of conflict pentagons in it, and that the number of replicas of each of these types is sufficiently large to guarantee that the conflict is universal (i.e. that it holds for a state described by an arbitrary real vector). We now proceed to demonstrate this point.\\

The top and bottom halves of Table \ref{tab:30-15} show one example of a division of the 60 states into two sets of 30 states that each give a parity proof of the KS theorem. Both these 30-state sets contain a large number of conflict pentagons, whose details are summarized in Table \ref{tab:Conflict}. Although both sets contain the same three types of conflict pentagons, their numbers in the two cases are different. The largest eigenvalue of any pentagon operator in these sets is 2.1778, which is smaller than the value of $2.236$ for a qutrit but larger than the value of $2.171$ found in the 18-ray set. However the more interesting point is that the presence of a large number of conflict pentagons in each of these sets suggests that the violation of the corresponding pentagon inequalities might be universal. To explore this, we parameterized an arbitrary real four-dimensional vector as $\overrightarrow{r}=(\cos\phi\sin\theta_{1}\sin\theta_{2},\sin\phi \sin\theta_{1} \sin\theta_{2}, \cos\theta_{1} \sin\theta_{2},\cos\theta_{2})$, with $0\leq \phi < 2\pi$ and $0\leq \theta_{1},\theta_{2} < 2\pi$, and calculated the quantities $V_{A}= \mathrm{Max} \langle\overrightarrow{r}\mid\Sigma_{A}\mid\overrightarrow{r}\rangle$ and $V_{B}= \mathrm{Max} \langle\overrightarrow{r}\mid\Sigma_{B}\mid\overrightarrow{r}\rangle$ as we varied over $\overrightarrow{r}$ over a fine mesh in projective Hilbert space, where by $V_{A}$ (or $V_{B}$ ) we mean the maximum calculated over all the pentagon operators in set A (or set B). We found that for any choice of $\overrightarrow{r}$, $V_{A} \geq 2.059$ and $V_{B} \geq 2.020$. In other words, we found that for an arbitrary state represented by a real four-dimensional vector, there is always at least one conflict pentagon in set A and one in set B that lead to violations of the pentagon inequality.\\

It should be stressed that the violation of the pentagon inequalities by our 30-ray sets is universal only for states described by real vectors. We are unaware of any other Kochen-Specker set of vectors that leads to a universal violation. It would, of course, be interesting to find an even smaller set of vectors that achieves this goal.

\begin{table}[ht]
\centering 
\begin{tabular}{|c c c c |c c c c |c c c c |} 
\hline 
1   &  2   &  3   &  4  &  14  &  60   & 34   &  1  &  48 &    5  &  32 & 58\\
\hline
13   & 14   & 15 &   16   & 13   & 32   & 50  &  41  &  19 &   25   &  6 & 50\\
\hline
41 &   42 &   43 &   44  &  25   & 44 &    2  &  53   & 34   & 19   & 48 & 54\\
\hline
31  &  42  &  51  &  16  &  58  &  36 &    15   &  4  &  46 &   31  &  60  & 6\\
\hline
43  &  54  &   3 &   28  &  36  &  53  &  20 &   46 &   20  &   5  &  51 & 28\\
\hline
 \multicolumn{12}{c}{} \\
\hline 
9   & 10  &  11  &  12   &  7  &  18  &  27 &   52  &   9  &  35  &  39 & 52\\
\hline
21  &  22  &  23  &  24  &  18 &   47  &  33  &  55  &  12 &   29   & 56 &  22\\
\hline
37  &  38 &   39  &  40  &  30 &   59  &  45  &   7 &   59  &  26  &  37  & 21\\
\hline
56  &  45   & 17 &   35  &  49  &   8  &  26  &  17 &   11   & 38  &  49  & 33\\
\hline
8  &  57  &  29  &  47 &   57 &   23  &  27  &  40 &   10  &  55 &   24 & 30\\
\hline 
\end{tabular}
\caption{The 30 states in the upper and lower tables have been arranged in 15 bases of four states each in such a way that each state occurs in exactly two of the bases, thus giving a ``parity proof" of the KS theorem. The 30 rays in the top table will be referred to as Set A, and the 30 in the bottom one as Set B.}
\label{tab:30-15} 
\end{table}

\begin{table}[ht]
\centering 
\begin{tabular}{||c | c | c c c c c ||} 
\hline\hline 
Max. Eigenvalue & Number & \multicolumn{5}{|c||}{States in Example} \\
\hline
2.1778 & 210 & 1 &  2 &  13 & 41 & 34  \\
2.1142 & 420 & 1  & 2 &  13 & 16 & 42 \\
2.0850 & 360 & 1 &  2 &  13 & 41 & 20 \\
\hline\hline 
 \multicolumn{7}{c}{} \\
 \hline\hline 
Max. Eigenvalue & Number & \multicolumn{5}{|c||}{States in Example} \\
\hline
2.1778  & 180 &  7  & 8  & 17 & 56 & 22 \\
2.1142  & 420 &  7  & 8  & 17 & 35  & 52 \\
2.0850  & 360 &  7  & 8  & 17 & 10 & 27 \\
\hline\hline 
\end{tabular}
\caption{Details of the conflict pentagons in the two 30-state sets of Table \ref{tab:30-15}, with the top table pertaining to Set A and the bottom one to Set B.  Both sets contain the same three types of conflict pentagons (with the maximum eigenvalues indicated), but their numbers in the two cases are different. An example of each of type of conflict pentagon in the two sets is indicated.}
\label{tab:Conflict} 
\end{table}

It has been pointed out in \cite{Bengtsson} that, in the so-called ``magic basis", any real four-dimensional vector represents a maximally entangled state and that any such state can be transformed into any other one by local operations on a pair of qubits. This remark suggests two alternative interpretations of the 60 real vectors in Table \ref{tab:Ray}. The first is to regard them as the states of a pair of qubits in the standard basis, in which case the states 1-8 are product states and the rest are entangled (but not maximally entangled). The second is to regard them as the states of  the qubits expressed in the ``magic basis", in which case all 60 vectors represent maximally entangled states. The latter interpretation has the merit that it allows all 60 states to be obtained from any basis set among them by purely local operations. This would make the task of measuring the  projectors in the pentagon operators much easier in an experimental test of the the pentagon inequalities based on these states. The local operations that can be used to transform states 1-4 into any of the other states are easily worked out, but we omit the details here.

\section{\label{sec:Ngons}N-gons in the 600-cell}

Let us define an N-gon as a set of N states that can be arranged at the vertices of an N-sided polygon in such a way that their Kochen-Specker diagram is represented by the N-gon with all its sides drawn in (and no other pairs of vertices connected). It is an interesting feature of the 600-cell that it has N-gons with every value of N from 5 to 15 in it, as shown in Table \ref{tab:Ngons}. The first row shows that the 600-cell has 22,320 different pentagons in it (of just the three types shown in Table \ref{tab:Conflict}), of which only 18,000 are conflict pentagons. These 18,000 pentagons are numerous enough to make the violation of the pentagon inequality universal, but using them all would amount to unnecessary ``overkill". Accordingly, we used only the pentagons in the two 30-ray sets of Table \ref{tab:30-15} to give a more economical demonstration of our result.\\

\begin{table}[ht]
\centering 
\begin{tabular}{|c | c | c || c |} 
\hline\hline 
& Number & Number of & Number in\\
N & of N-gons & Conflict N-gons & A \& B Sets\\
\hline
5 & 22,320 & 18,000 & 1,200\\
6 & 94,200 & 0 & 2,100 \\
7 & 302,400 & 14,400 & 3,030\\
8 & 432,000 & 0 & 1,110\\
9 & 436,800 & 0 & 630 \\
10 & 862,560 & 0 & 0\\
11 & 410,400 & 0 & 0\\
12 & 175,800 & 0 & 0\\
13 & 302,400 & 0 & 0\\
14 & 43,200 & 0 & 0\\
15 & 33,120 & 0 & 8\\
\hline\hline 
\end{tabular}
\caption{N-gons in the 600-cell. For each value of N, the second and third columns show the total number of N-gons as well as the (smaller) number of conflict N-gons in all 60 states, while the fourth column shows the total number of N-gons in each of the 30-state sets A and B. The number of conflict pentagons and heptagons in sets A and B is indicated in Tables \ref{tab:Conflict} and \ref{tab:Heptagons}.}
\label{tab:Ngons} 
\end{table}

An argument similar to the one for $N = 5$ shows that for $N > 5$ a violation of noncontextual realism requires that $\langle\Sigma\rangle > N/2$ if $N$ is even or $\langle\Sigma\rangle > (N-1)/2$ if $N$ is odd, with the N-gon operator now being the sum of the N projectors involved. We find that the only higher N-gons that lead to a conflict with noncontextual realism are heptagons (or 7-gons). The conflict is very weak, with $\langle\Sigma\rangle$ only being slightly larger than 3, as shown in Table \ref{tab:Heptagons}. Neither the heptagons in sets A or B suffice to establish a universal violation, but that is not a drawback since the pentagons already achieved that goal.\\

\begin{table}[ht]
\centering 
\begin{tabular}{||c | c | c c c c c c c||} 
\hline\hline 
Max. Eigenvalue & Number & \multicolumn{7}{|c||}{Rays in Example}\\
\hline
3.005 & 120 & 1 &  2 &  13 & 32 & 58 & 36 & 34 \\
\hline\hline 
\multicolumn{9}{c}{} \\
\hline\hline 
Max. Eigenvalue & Number & \multicolumn{7}{|c||}{Rays in Example} \\
\hline
3.043 & 120 & 7 &  8 &  17 & 10 & 37 & 21 & 22 \\
3.005 & 60 & 7 &  8 & 17 & 35 & 33 & 23 & 22 \\
\hline\hline 
\end{tabular}
\caption{The top table shows the only type of conflict heptagon in set A, while the bottom one shows the two different types that exist in set B. This table, like Table \ref{tab:Conflict}, illustrates the geometrically distinct characters of sets A and B.}
\label{tab:Heptagons} 
\end{table}

Although the N-gons with $N > 7$ do not lead to a conflict with nonlocal realism of the sort considered here, it is possible that they could be used as scaffolds for other constructions that do. We thought it worth pointing out their existence in case they find this or other applications.\\

In conclusion, we have demonstrated that the system of 60 states derived from a 600-cell yields a variety of pentagon inequalities that can be used as the basis for an experimental disproof of noncontextual realism that is universal (in the sense that it holds for all states represented by real four-dimensional vectors). It is interesting that the same 30-ray sets that provide parity proofs of the KS theorem can also be exploited to give this alternative disproof of noncontextual realism.


\begin{thebibliography}{10}
\providecommand{\url}[1]{{#1}}
\providecommand{\urlprefix}{URL }
\expandafter\ifx\csname urlstyle\endcsname\relax
  \providecommand{\doi}[1]{DOI \discretionary{}{}{}#1}\else
  \providecommand{\doi}{DOI \discretionary{}{}{}\begingroup
  \urlstyle{rm}\Url}\fi


\bibitem{Specker}
E.~Specker, in C.A.Hooker (Ed.), {\em{The Logico-Algebraic Approach to Quantum Mechanics. Volume 1: Historical Evolution}}, (Reidel, Dordrecht, 1975), p.135-140.

\bibitem{Klyachko}
A.A.~Klyachko, M.A. Can, S. Binicio\u{g}lu and A.S. Shumovsky: {\it Phys. Rev. Lett.} \textbf{{\bf 101}}, 020403 (2008).

\bibitem{Bengtsson}
P.~Badzi{\c a}g, I. Bengtsson, A. Cabello, H.H. Granstr\"om and J.-\AA. Larsson: {\it Found. Phys.} \textbf{{\bf 41}}, 414 (2011).

\bibitem{Liang}
Y.C.~Liang, R.W. Spekkens, H.M. Wiseman: {\it Phys. Rep.} \textbf{{\bf 506}}, 1 (2011).

\bibitem{Severini}
A.~Cabello, S. Severini and A. Winter: {\it {A}r{X}iv:1010.2163} (2010).

\bibitem{Cunha}
A.~Cabello and M.~Terra Cunha: Phys. Rev. Lett. \textbf{{\bf 106}}, 190401 (2011).

\bibitem{Aolita}
L.~Aolita, R. Gallego, A. Acin, A. Chiuri, G. Vallone, P. Mataloni and A. Cabello: {\it {A}r{X}iv:1105.3598} (2011).

\bibitem{Sadiq}
M.~Sadiq, P. Badzi{\c a}g, M. Bourennane and A. Cabello: {\it {A}r{X}iv:1106.4754} (2011).

\bibitem{Kirchmair}
G.~Kirchmair, F.~Z{\"a}hringer, R.~Gerritsma, M.~Kleinmann, O.~G{\"u}hne,
  A.~Cabello, R.~Blatt, C.F.~Roos: {\it Nature} \textbf{{\bf 460}}, 494 (2009).

\bibitem{Bartosik}
H.~Bartosik, J.~Klep, C.~Schmitzer, S.~Sponar, A.~Cabello, H.~Rauch, Y.~Hasegawa: {\it Phys. Rev. Lett.} \textbf{{\bf 103}}, 040403 (2009).

\bibitem{Amselem}
E.~Amselem, M.~R{\aa}dmark, M.~Bourennane, A.~Cabello: {\it Phys. Rev. Lett.} \textbf{{\bf 103}}, 160405 (2009).

\bibitem{Moussa}
O.~Moussa, C.A. Ryan, D.G. Cory, R.~Laflamme: {\it Phys. Rev. Lett.} \textbf{{\bf 104}}, 160501 (2010).

\bibitem{Waegell2011}
M.Waegell, P.K.Aravind, N.D.Megill and M.~Pavi{\v c}i{\'c}: {\it Found. Phys.} \textbf{{\bf 41}}, 883 (2011).

\bibitem{KS1967}
S.~Kochen, E.P.~Specker: {\it J. Math. Mech.} \textbf{{\bf 17}}, 59 (1967). See also J.S.~Bell: {\it Rev. Mod. Phys.} \textbf{{\bf 38}}, 447 (1966).
\newblock Reprinted in J.S.~Bell: {\em{Speakable and Unspeakable in Quantum Mechanics}}, (Cambridge University Press, Cambridge, 1987).

\bibitem{noncon}
The assumption of noncontextuality implies specifically that the value obtained for a projector is the same no matter which of the two orthogonal projectors it is measured together with.

\bibitem{Cabello1996}
A.~Cabello, J.M.~Estebaranz, G.~{Garc{\'\i}a-Alcaine}: {\it Phys. Lett. A} \textbf{{\bf 212}}, 183 (1996).

\bibitem{Peres1991}
A.~Peres: {\it J. Phys. A} \textbf{{\bf 24}}, L175 (1991).

\end{thebibliography}

\end{document}